\documentclass[11pt]{article}
\usepackage{amsmath, amscd}
\usepackage{amssymb}
\usepackage{latexsym}

\catcode`@=11
\renewcommand{\section}{\@startsection{section}{1}{0pt}{\medskipamount}
{\medskipamount}{\large\bf}}
\numberwithin{equation}{section}
\catcode`@=12
\def\a{\alpha}
\def\b{\beta}
\def\ga{\gamma}

\def\de{\delta}
\def\eps{\epsilon}
\def\ve{\varepsilon}

\def\Si{\Sigma}

\def\th{\theta}
\def\om{\omega}

\newcommand{\R}{\mathbb R}

\newcommand{\Jcal}{{\cal J}}
\newcommand{\Gcal}{{\cal G}}
\newcommand{\Acal}{{\cal A}}

\newcommand{\Mcal}{{\cal M}}
\newcommand{\Fcal}{{\cal F}}
\newcommand{\Ncal}{{\cal N}}

\newcommand{\Ocal}{{\cal O}}

\newcommand{\C}{{\mathbb C}}

\newcommand{\gfrak}{{\mathfrak g}}
\newcommand{\mfrak}{{\mathfrak m}}
\newcommand{\hfrak}{{\mathfrak h}}

\def\im{\textrm{i}}
\def\tr{\textrm{tr}}

\def\Lie{\textrm{Lie}}
\def\diff{\textrm{d}}
\def\pa{\mbox{$\partial$}}
\def\sfrac#1#2{{\textstyle\frac{#1}{#2}}}
\def\+{\dagger}
\def\={\ =\ }
\def\<{\langle}
\def\>{\rangle}
\def\und{\qquad\textrm{and}\qquad}
\def\and{\quad\textrm{and}\quad}
\def\with{\quad\textrm{with}\quad}
\def\for{\quad\textrm{for}\quad}
\def\Id{\mathrm{Id}}
\def\1{\bar 1}
\def\2{\bar 2}
\def\3{\bar 3}
\def\4{\bar 4}

\begin{document}

\begin{titlepage}
\setcounter{page}{0}
\begin{flushright}
ITP-UH-18/16
\end{flushright}

\hspace{2.0cm}

\begin{center}

{\Large\bf Superstring limit of Yang-Mills theories }

\vspace{12mm}

{\large
Olaf Lechtenfeld and Alexander D. Popov
}\\[8mm]

\noindent {\em
Institut f\"ur Theoretische Physik {\rm and} Riemann Center for Geometry and Physics \\
Leibniz Universit\"at Hannover \\
Appelstra\ss e 2, 30167 Hannover, Germany }\\[4mm]
{ Email: lechtenf@itp.uni-hannover.de, popov@itp.uni-hannover.de
}\\[6mm]

\vspace{10mm}

\begin{abstract}
\noindent 
It was pointed out by Shifman and Yung that the critical superstring on $X^{10}=\R^4\times Y^6$, where  $Y^6$
is the resolved conifold, appears as an effective theory for a U(2) Yang-Mills-Higgs system with four fundamental
Higgs scalars defined on $\Sigma_2\times \R^2$, where $\Sigma_2$ is a two-dimensional Lorentzian manifold. 
Their Yang-Mills model supports semilocal vortices on $\R^2\subset\Sigma_2\times \R^2$ with a moduli space~$X^{10}$. 
When the moduli of slowly moving thin vortices depend on the coordinates of~$\Sigma_2$, the vortex strings 
can be identified with critical fundamental strings. We show that similar results can be obtained for the low-energy
limit of pure Yang-Mills theory on $\Sigma_2\times T^2_p$, where $T^2_p$ is a two-dimensional torus with a puncture~$p$. 
The solitonic vortices of Shifman and Yung then get replaced by flat connections.
Various ten-dimensional superstring target spaces can be obtained as moduli spaces of flat connections on $T^2_p$,
depending on the choice of the gauge group. The full Green-Schwarz sigma model requires extending the gauge group
to a supergroup and augmenting the action with a topological term.
\end{abstract}

\end{center}
\end{titlepage}

\section {Introduction}

\noindent
Recently, Koroteev, Shifman and Yung~\cite{1, 2, 3} have shown that U(2) solitonic vortex strings
in certain $\Ncal{=}\,2$ super-Yang-Mills theories have an effective infrared dynamics of a critical fundamental string
on a ten-dimensional target space $X^{10}=\R^4\times Y^6$, where $Y^6$ is the resolved conifold.\footnote{
For fine survey articles on non-Abelian vortices, their moduli spaces and reductions to effective
$d{=}2$ sigma models see e.g.~\cite{4,5,6,7} and references therein.}
More precisely, $\Ncal{=}\,2$ supersymmetric U(2) Yang-Mills-Higgs theory on $\Sigma_2\times \R^2$, 
where $\Si_2$ is a two-dimensional Lorentzian manifold, with a Fayet-Illiopoulos term and four flavor hypermultiplets in 
the fundamental of~U(2) admits non-Abelian semilocal vortices on~$\R^2$ whose (translational, orientational and size) moduli 
are parametrized by~$X^{10}$. Allowing the vortex moduli to depend on the coordinates of $\Si_2$ yields a string sigma model
with worldsheet $\Si_2$ and target~$X^{10}$, which describes the effective vortex dynamics.

In~\cite{1, 2, 3} the $\Ncal{=}\,2$ super-Yang-Mills model with fundamental matter was chosen because it admits vortex solutions
with a Ricci-flat ten-dimensional moduli space. Also the metric on~$\Si_2$ was taken as an independent variable.
These two assumptions differ from earlier treatments~\cite{8,9}, where $\Ncal{=}\,4$ super-Yang-Mills theory on 
$\Si_2\times\tilde\Si_2$ in the infrared limit ($\tilde\Si_2$ shrinking to a point) was reduced to certain sigma models on 
$\Si_2$ whose target space is the moduli space~$\Mcal$ of flat connections\footnote{
{}From twisted super-Yang-Mills theories one can also get the cotangent bundle $T^*\Mcal$ as target space,
see e.g.~\cite{9,10}.}
on a Riemann surface~$\tilde\Si_2$.

In pure Yang-Mills theory and its standard supersymmetric extensions one gets flat connections instead of vortices.
This is just as well, as we will demonstrate for $\tilde\Si_2=T^2_p$, a two-dimensional torus $T^2$ with a puncture~$p$. 
This case is simpler than that of a circle $S^1$ or a disk $H^2$ considered earlier~\cite{11, 12, 13}, 
and it deserves a separate study. Therefore, in this paper we investigate the infrared limit of pure Yang-Mills theory
on~$\Si_2\times T^2_p$, and we describe further examples of string backgrounds which can be obtained in this framework.

The organization of this paper is as follows. 
In Section~2 we describe a four-manifold $M^4=\Si_2\times T^2_p$ with an $\ve$-deformed metric and introduce
the $\ve$-dependent Yang-Mills action on $M^4$ with a gauge group $G$, where $\ve\in [0,\infty )$. 
In Section~3 we perform the low-energy limit $\ve\to 0$ under which the Yang-Mills theory reduces to a stringy sigma model. 
We explain in some detail how gauge-field moduli become coordinates on the sigma-model target space (cf.~\cite{14, 9, 15}). 
Its effective action and Virasoro-type constraints will be derived. 
In Section~4 we provide a number of examples of the above-mentioned target spaces, 
including supercosets such as PSU(2,2$|$4)/SO(4,1)$\times$SO(5) related with AdS$_5\times S^5$. 
The Conclusions summarize our findings and point out possible generalizations and applications.

\section{Yang-Mills theory}

\noindent {\bf Lie (super)groups.}  
In our approach the Green-Schwarz superstring action can be obtained from Yang-Mills theory in
four dimensions with a supergroups as structure group (cf.~\cite{11, 12, 13}). 
However, here we mainly restrict ourselves to deriving the bosonic part of superstring actions, 
similarly as in~\cite{1,2,3}. This will make the discussion simpler and clearer.
Green-Schwarz actions for various target spaces and the corresponding Lie supergroups will be briefly discussed in Section~4.

For the Yang-Mills structure group we consider a Lie group $G$ with a closed subgroup $H$. Then, for the
Lie algebras  $\gfrak =\,$Lie$\,G$ and  $\hfrak =\,$Lie$\,H$ we have\footnote{
This splitting will be used later in defining a boundary condition for gauge connections.}
\begin{equation}\label{1}
\gfrak=\hfrak\oplus\mfrak\ ,
\end{equation}
where $\mfrak$ is the orthogonal complement of $\hfrak$ in $\gfrak$ with respect to a metric $\<\;,\,\>$ on $\gfrak$.
For matrix (super)algebras, $\<X,Y\>=\textrm{(S)}\tr(XY)$ is the ordinary trace or supertrace.
For additive groups like $\R^k$, it denotes the ordinary metric on vector spaces.

\medskip

\noindent {\bf Gauge fields.} We consider Yang-Mills theory on a direct product manifold 
\begin{equation}
M^4=\Si_2\times T_p^2 \qquad\textrm{with coordinates}\quad 
(x^\mu )=(x^a, x^i) \quad\textrm{for}\quad a=1,2 \and i=3,4\ ,
\end{equation}
where
$\Si_2$ is a two-dimensional Lorentzian manifold with a metric tensor $g^{}_{\Si_2}=(g_{ab})$, and
$T^2_p=T^2\setminus\{p\}$ is a two-dimensional torus with a point $p$ removed (the puncture) and a
metric $g^{}_{T^2}=(g_{ij})$. We will just write $T^2$ (omitting the puncture) since we do not consider 
other tori in this paper. Then the metric tensor on $M^4$ reads $(g_{\mu\nu})= (g_{ab}, g_{ij} )$ with
$\mu,\nu = 1,\ldots,4$. Fixing momentarily the size of~$T^2$, det$(g_{ij})=1$, the metric $g^{}_{T^2}$ 
still depends on the complex shape parameter~$\tau$. For simplicity we choose the square torus $\tau=\im$, 
i.e.~we identify $x^i\sim x^i+1$ for both homology circles.

We consider a topologically trivial bundle over $M^4$ (the principal $G$-bundle $P$ and an associated vector bundle
$E\to M^4$) with a gauge potential $\Acal =\Acal_{\mu}\diff x^\mu$ (a connection) taking values in $\gfrak$.
The $\gfrak$-valued gauge field (the curvature) reads
\begin{equation}\label{2}
 \Fcal =\sfrac12\Fcal_{\mu\nu}\diff x^\mu \wedge \diff x^\nu\with \Fcal_{\mu\nu} =\partial_\mu\Acal_\nu -
\partial_\nu\Acal_\mu
 + [\Acal_\mu , \Acal_\nu]\ .
\end{equation}
On $M^4=\Si_2\times T^2$ we have the obvious splitting
\begin{equation}\label{3}
 \diff s^2 = g_{\mu\nu}\diff x^\mu \diff x^\nu = g_{ab}\diff x^a \diff x^b + g_{ij}\diff x^i \diff x^j\ ,
\end{equation}
\begin{equation}\label{4}
\Acal =\Acal_{\mu}\diff x^\mu=\Acal^{}_{\Si_2} + \Acal^{}_{T^2}= \Acal_{a}\diff x^a+\Acal_{i}\diff x^i\ ,
\end{equation}
\begin{equation}\label{5}
 \Fcal =\sfrac12\Fcal_{ab}\diff x^a \wedge \diff x^b + \Fcal_{ai}\diff x^a \wedge \diff x^i
+\sfrac12\Fcal_{ij}\diff x^i \wedge \diff x^j\ .
 \end{equation}
We note that there are mixed components $\Fcal_{ai}$ in (\ref{5}).

Let us now deform the metric (\ref{3}) and introduce
\begin{equation}\label{6}
 \diff s^2_\ve = g_{\mu\nu}^\ve\,\diff x^\mu \diff x^\nu = g_{ab}\diff x^a \diff x^b + \ve^2 g_{ij}\diff x^i \diff x^j
 \qquad\textrm{hence}\quad g^\ve_{ab}= g_{ab} \and g^\ve_{ij}= \ve^2 g_{ij}\ ,
\end{equation}
where $\ve\in [0, \infty )$ is a dimensionless real parameter. Then $\det (g_{\mu\nu}^\ve )=\ve^4\det (g_{ab})$ and
\begin{equation}\label{7}
 \Fcal^{ab}_\ve = g_\ve^{ac}g_\ve^{bd}\Fcal_{cd}= \Fcal^{ab}\ ,\quad \Fcal^{ai}_\ve = g_\ve^{ac}g_\ve^{ij}\Fcal_{cj}=
 \ve^{-2}\Fcal^{ai}\ ,\quad
 \Fcal^{ij}_\ve = g_\ve^{ik}g_\ve^{jl}\Fcal_{kl}=\ve^{-4}\Fcal^{ij}\ ,
\end{equation}
where the indices in $\Fcal^{\mu\nu}$ are raised by the nondeformed metric tensor~$g^{\mu\nu}$. 
One can introduce on $T^2$ adapted coordinates $y^i=\ve x^i$ for which $y^i\sim y^i+\ve$. 
In other words, the deformation reintroduces the size modulus of $T^2$: 
for $\ve^2\to 0$ the torus shrinks to a point.\footnote{
It is usually assumed that $\Acal_\mu$ and $\Fcal_{\mu\nu}$ smoothly depend on $\ve^2$ 
with a well-defined limit for $\ve^2\to 0$.} 
This limit is equivalent to the low-energy limit of gauge theory on $\Si_2\times T^2$~\cite{8}.

\medskip

\noindent {\bf Yang-Mills action.} For the deformed metric  (\ref{6}) the Yang-Mills action functional is
\begin{equation}\label{8}
S_\ve=\int_{M^4} \diff^4x\,\sqrt{|\det g^{}_{\Si_2}|}\,\left\{\ve^2\langle\Fcal_{ab}, \Fcal^{ab}\rangle 
+ 2\langle\Fcal_{ai}, \Fcal^{ai}\rangle + \ve^{-2}\langle\Fcal_{ij}, \Fcal^{ij}\rangle\right\}\ .
\end{equation}
For $\ve^2=1$ one has the standard Yang-Mills Lagrangian on $M^4=\Si_2\times T^2$ with the nondeformed metric
(\ref{3}), and for $\ve^2\to 0$ it reduces to a stringy sigma-model action on $\Si_2$ as we will see in a moment.

We play with the metric on $T^2$, but the metric on $\Si_2$ can be dynamical, 
i.e.~the Yang-Mills model is coupled to (two-dimensional) gravity.
Therefore, one can add to the Lagrangian in (\ref{8}) the term
\begin{equation}\label{9}
\sqrt{|\det g^{}_{M^4_{\ve}}|}\,R^{}_{M^4_{\ve}}\=\ve^2\sqrt{|\det g^{}_{\Si_2}|}\,R^{}_{\Si_2}\ ,
\end{equation}
where $R^{}_{M^4_{\ve}}$ and $R^{}_{\Si_2}$ are the scalar curvatures of $M^4$ and of  $\Si_2$, respectively,
with the metric (\ref{6}). 
The term (\ref{9}) does not contribute to the equations of motion since integration of (\ref{9}) over $M^4$ gives
a topological invariant of ${\Si_2}$. This is not so if we couple (\ref{9}) with the dilaton field $\Phi$, but anyway the term
(\ref{9}) vanishes in the limit $\ve^2\to 0$ which we consider. For this reason we do not add (\ref{9}) to the Yang-Mills
Lagrangian in (\ref{8}).

The Yang-Mills equations following from  (\ref{8}) are
\begin{equation}\label{10}
\ve^2 D_a\Fcal^{ab} +  D_i\Fcal^{ib} \=0\ ,
\end{equation}
\begin{equation}\label{11}
D_a\Fcal^{aj} +  \ve^{-2} D_i\Fcal^{ij} \=0\ ,
\end{equation}
where $D_a, D_i$ are Yang-Mills covariant derivatives on the curved background $M^4=\Si_2\times T^2$.
The Euler-Lagrange equations for $g^{}_{\Si_2}$ yield the constraint equations 
\begin{equation}\label{12}
T_{ab}^{\ve}\ \equiv\ \ve^2\bigl(g^{cd}\langle\Fcal_{ac},\Fcal_{bd}\rangle{-}
\sfrac14 g_{ab}\langle\Fcal_{cd},\Fcal^{cd}\rangle\bigr){+}g^{ij} \langle\Fcal_{ai},\Fcal_{bj}\rangle{-}\sfrac12
 g_{ab}\langle\Fcal_{ci},\Fcal^{ci}\rangle{-}\sfrac14 \ve^{-2}g_{ab}\langle\Fcal_{ij},\Fcal^{ij}\rangle\=0
\end{equation}
for the Yang-Mills energy-momentum tensor $T_{\mu\nu}^{\ve}$, i.e.~its components along $\Si_2$ are vanishing.
Its other components, $T_{ij}^{\ve}$ or $T_{aj}^{\ve}$, are not constrained.
Note that we might employ the invariance under diffeomorphisms on $\Si_2$ to locally fix its metric, 
e.g., to a flat metric in the conformal gauge. Nevertheless, (\ref{12}) must be added as an external constraint.

\section{Low-energy effective action}

\noindent {\bf Adiabatic limit.} As usual in the adiabatic approach (see e.g.~\cite{16,17}), we assume that
the connection $\Acal$ for small $\ve^2$ can be expanded in a Taylor series in $\ve^2$, i.e.
$\Acal =\Acal^0+\ve^2\Acal^1 + O(\ve^4)$. In particular, $\Acal^{}_{T^2} =\Acal^0_{T^2}
+\ve^2\Acal^1_{T^2} +  O(\ve^4)$ and therefore
\begin{equation}\label{3.1}
\Fcal_{ij}\=\Fcal_{ij}^0 +\ve^2(D^0_i\Acal^1_j-D^0_j\Acal^1_i) + O(\ve^4)\ ,
\end{equation}
where $D^0_i=\partial_i +[\Acal^0_i, \cdot ]$ and $\Fcal_{ij}^0=[D^0_i,D^0_j]$. From (\ref{8}) one sees that the term
$\ve^{-2}\langle\Fcal_{ij}^0,\Fcal^{0\,ij}\rangle$ in the Yang-Mills action diverges when  $\ve^2\to 0$. To
avoid this one should impose the condition
\begin{equation}\label{3.2}
 \Fcal_{ij}^0=0
\end{equation}
on the components of the Yang-Mills field along $T^2$. We denote by $\Mcal^{}_{T^2}$ the moduli space of solutions
(flat connections) to the equations (\ref{3.2}) on $T^2$ with a puncture at $p$. It is known  (see e.g.~\cite{17,18}) that
terms of order $\ve^{2k}$ in $\Acal^{}_{T^2}$ for $k\ge 1$ are orthogonal to $\Mcal^{}_{T^2}$ and
yield some massive modes in the effective theory on $\Si_2$. A consideration of these modes goes beyond the scope of
this paper. In the limit $\ve^2\to 0$ we keep only $\Acal^0$ and $\Fcal^0$ (zero-mode moduli approximation),
and from now on we omit the index ``0'' in connection  $\Acal^0$ and the curvature $\Fcal^0$.

In the adiabatic approximation (when  $\ve^2\to 0$), the Yang-Mills action (\ref{8}) becomes
\begin{equation}\label{3.3}
S_0\ =\int_{M^4} \diff^4x\,\sqrt{|\det g^{}_{\Si_2}|}\,\langle\Fcal_{ai},\Fcal^{ai}\rangle\ .
\end{equation}
As equations of motion one gets
\begin{equation}\label{3.4}
D_i\Fcal^{ib}\ \equiv\ 
\frac{1}{\sqrt{|\det g^{}_{\Si_2}|}}\,\partial_i\left(\sqrt{|\det g^{}_{\Si_2}|}\,g^{ab}g^{ij}\Fcal_{aj}\right)
+ [\Acal_i , \Fcal^{ib}]\=0\ ,
 \end{equation}
\begin{equation}\label{3.5}
D_a\Fcal^{aj}\ \equiv\ 
\frac{1}{\sqrt{|\det g^{}_{\Si_2}|}}\,\partial_a\left(\sqrt{|\det g^{}_{\Si_2}|}\,g^{ab}g^{ij}\Fcal_{ib}\right)
+ [\Acal_a , \Fcal^{aj}]\=0\ .
 \end{equation}
The constraint equations (\ref{12}) in the limit $\ve^2\to 0$ have the form
\begin{equation}\label{3.6}
T^0_{ab}\ \equiv\ g^{ij} \langle\Fcal_{ai},\Fcal_{bj}\rangle -\sfrac12
 g_{ab}\langle\Fcal_{ci},\Fcal^{ci}\rangle \=0\ .
\end{equation}

\medskip

\noindent {\bf Flat connection on $T^2$.} It is well known that on smooth tori $T^2$
(compact, without punctures) there are no irreducible flat connections $\Acal^{}_{T^2}\in \gfrak$~\cite{19}.
There exist only reducible flat connections which are constant and take values in the Cartan subalgebra
of $\gfrak$ (see e.g.~\cite{8}). This so-called ``abelianization" theorem is widely used in the literature
on Yang-Mills confinement on $\R^3\times S^1$ and $\R^2\times T^2$. However, this theorem
is not valid on Riemann surfaces with punctures or fixed points (see e.g.~\cite{20,21,22}). In particular, on
tori $T^2$ with a puncture one can find irreducible flat connections on $G$-bundles over $T^2$~\cite{21}, and
the same is true for higher genus (see e.g.~\cite{21,22}).\footnote{Irreducible flat connections on complex
vector bundles over smooth Riemann surfaces define {\it stable} holomorphic bundles. For vector bundles over Riemann
surfaces with punctures, stability is replaced with Seshadri's notion of parabolic stability~\cite{20,21}.}

Flat connections, i.e.~solutions of (\ref{3.2}), on a torus $T^2$ with a puncture can be described as
follows~\cite{21}. 
The puncture $p\in T^2$ can be considered as infinity similar to the north pole on the two-sphere $S^2$, 
and one can introduce cylindrical coordinates $(\varrho, \th)$ on a small disk centered at $p$ via
$x^3=\exp(-\varrho)\cos\th$ and $x^4=\exp(-\varrho)\sin\th$.
The group of gauge transformations is defined as the Banach Lie group
\begin{equation}\label{3.7}
\Gcal^{}_{T^2}\= \bigl\{ \,\textrm{smooth maps}\,g: \ T^2\to G\,\bigr\}\ ,
\end{equation}
whose topology is described in~\cite{21,22}.
On the flat connections $\Acal^{}_{T^2}$ we impose the boundary condition
\begin{equation}\label{3.8}
 \Acal^{}_{T^2}=\Acal_\varrho\diff\varrho + \Acal_\th\diff\th\quad\to\quad 
 \Acal_p=a\,\diff\th\quad\for \varrho\to\infty\ .
\end{equation}
Here $a$ is either an arbitrary element of $\mfrak$ for the decomposition $\gfrak=\hfrak\oplus\mfrak$ introduced in (\ref{1}),
or $a=g_p h_0 g_p^{-1}$, where $h_0\in\hfrak$ is fixed and $g_p\in G/H$ is arbitrary.
Then $\Acal_p$ is parametrized by $g_0=\exp(2\pi a)\in G/H$ for $a\in\mfrak$ or $g_p\in G/H$, 
where the case $H=\{\Id\}$ is included. 
If we denote by $\Ncal$ the space of all such flat connections then their moduli space is
\begin{equation}\label{3.9}
 \Mcal \= \Ncal/ \Gcal^{}_{T^2}\= G/H \ .
\end{equation}
In other words, the gauge group (\ref{3.7}) forms the fibres over points in $\Mcal$ for the bundle
\begin{equation}\label{3.11}
\begin{CD}
\pi :\quad \Ncal   @>{\Gcal^{}_{T^2}}>>  \Mcal = G/H\ .
\end{CD}
\end{equation}
Note that, if $G/H$ is a adjoint orbit, e.g.~the K\"ahler coset space
\begin{equation}\label{3.11a}
G/H = \mbox{U}(N)/  \mbox{U}(N_1)\times ...\times \mbox{U}(N_k) \with N_1+...+N_k=N \ ,
\end{equation}
then $\Mcal$ is the moduli space of irreducible flat connections on vector bundles with parabolic structure
(see~\cite{20, 21} for more details).

\medskip

\noindent {\bf Moduli space. } We endow our moduli space $\Mcal$ of flat connections on a punctured $T^2$ with 
local coordinates~$(\phi^\a)$, with $\a=1,\ldots,\textrm{dim}(\Mcal)$. In the adiabatic
approach, the moduli approximation assumes that $\phi^\a$ depend on $x^a\in \Si_2$~[14,4--10]
In this way, the moduli of flat connections on $T^2$ define a map
\begin{equation}\label{3.12}
\phi :\quad \Si_2\ {\to}\ \Mcal \qquad\textrm{via}\quad (x^a)\mapsto \{\phi^\a (x^a)\}
\qquad\textrm{so that}\quad \Acal^{}_{T^2} =\Acal^{}_{T^2} (\phi^\a (x^a), x^i)\ .
\end{equation}
Now our space $\Ncal$ of solutions to (\ref{3.2}) depends on $x\in\Si_2$ as well as on elements $g$ of the gauge group
$\Gcal^{}_{T^2}$. In fact, for any fixed $x\in \Si_2$ and $\Gcal^{}_{x}=\Gcal^{}_{T^2}(x^a)$,
the gauge group $\Gcal$ of the full theory on $M^4=\Si_2\times T^2$ coincides with $\Gcal^{}_{T^2}$. 
Said differently, for any fixed $x\in \Si_2$ we have a copy of the moduli space $\Mcal_x=\Ncal_x/ \Gcal_x \cong G/H$ 
of flat connections on $T^2$.

The maps (\ref{3.12}) are not arbitrary -- they are constrained by the equations (\ref{3.4})-(\ref{3.6}). Since
$\Acal^{}_{T^2}$ is a flat connection on $T^2$ for {\it any\/} point in $\Si_2$, the derivatives $\pa_a\Acal_i$ have to satisfy
the linearized (around $\Acal_i$) flatness equations (\ref{3.2}). In other words, $\pa_a\Acal_i$ belong to the
tangent space $T^{}_{\Acal}\Ncal$ of the solution space $\Ncal$. Using the projection (\ref{3.11}), 
one can orthogonally decompose $\pa_a\Acal_i$ into two parts,
\begin{equation}\label{3.13}
T^{}_\Acal \Ncal\= \pi^*T^{}_\Acal \Mcal\oplus T^{}_\Acal \Gcal \qquad\Rightarrow
\end{equation}
\begin{equation}\label{3.14}
\frac{\pa\Acal_i}{\pa\phi^\a}=\xi_{\a i} + D_i\eps_\a \qquad\Leftrightarrow\qquad 
\pa_a\Acal_i =\frac{\pa\phi^\a}{\pa x^a}
\frac{\pa\Acal_i}{\pa\phi^\a}= (\pa_a \phi^\a)\xi_{\a i} + D_i\eps_a\ ,
\end{equation}
where
\begin{equation}\label{3.15}
\diff x^i D_i\eps_\a \in T^{}_{\Acal} \Gcal \und
\eps_a : =(\pa_a\phi^\a)\eps_\a \ ,
\end{equation}
i.e.~$\eps_\a$ are $\gfrak$-valued gauge parameters from the viewpoint of gauge theory on $T^2$, 
and $\xi_{\a}=\xi_{\a i}\diff x^i \in T^{}_\Acal \Mcal$ can be identified with vector fields on $\Mcal =G/H$.
Thus, $\xi_{\a}$ correspond to generators from the subspace $\mfrak$ in the Lie-algebra decomposition
$\gfrak =\hfrak \oplus \mfrak$.

The fields $\eps_a$ are determined by the gauge-fixing conditions
\begin{equation}\label{3.16}
g^{ij} D_i\xi^{}_{\a j}=0\qquad\Rightarrow\qquad g^{ij} D_iD_j\eps_a=g^{ij} D_i \pa_a\Acal_j \ .
\end{equation}
Note that, due to (\ref{3.2}), one can solve the first equation,
\begin{equation}
2\ve^{ij}D_iD_j = \ve^{ij}\Fcal_{ij}=0 \qquad\Rightarrow\qquad \xi_{\a j}=\ve_{j}^k D_k\xi_\a\ ,
\end{equation}
with $\ve_{j}^k := g^{kl}\ve_{lj}$ and $\ve^{ij}= g^{ik} \ve^{j}_k$.

\medskip

\noindent {\bf Effective action. } 
Recall that $\Acal_i$ obey (\ref{3.2}) and have the moduli space $\Mcal =G/H$, 
which parametrizes the boundary values of the connection at the puncture $p\in T^2$.
The case $H=\{\Id\}$ of a group manifold $\Mcal = G$ is included. 
On the other hand, the components $\Acal_a$ are yet free. 
It is natural to identify them with $\eps_a$~\cite{9,15},\footnote{
In fact, $\eps_\a\diff\phi^\a$ in (\ref{3.14}) is a connection on a $G$-bundle over $\Mcal$,
and $\eps_a\diff x^a$ from (\ref{3.15}) is the pull-back of the connection $\eps_\a\diff\phi^\a$ from the  $G$-bundle over
$\Mcal$ to the $G$-bundle over $\Si_2$~\cite{9}. Therefore, $\Acal_a\diff x^a$ and $\eps_a\diff x^a$ are connections on the
same bundle over $\Si_2$, and it is natural to identify them.} 
\begin{equation}\label{3.17}
\Acal_a=\eps_a \qquad\Rightarrow\qquad
\Fcal_{ai}\=\pa_a\Acal_i - D_i\Acal_a \= (\pa_a\phi^{\a})\xi_{\a i}\ \in T^{}_\Acal \Mcal\ .
\end{equation}
Substituting  (\ref{3.17}) into (\ref{3.4}), we see that  (\ref{3.4}) is resolved due to  (\ref{3.16}).
Plugging  (\ref{3.17}) into the action (\ref{3.3}), we get the effective sigma-model action
\begin{equation}\label{3.18}
S_0\ =\int_{\Si_2} \diff x^1 \diff x^2\,\sqrt{|\det g^{}_{\Si_2}|}\, g^{ab}\,\pa_a\phi^\a\,\pa_b\phi^\b\,G_{\a\b}\ ,
\end{equation}
where
\begin{equation}\label{3.19}
G_{\a\b}(\phi,\tau )\ = \int_{T^2} \diff x^3 \diff x^4\, g^{ij}\,\langle\xi_{\a i},\xi_{\b j}\rangle
\end{equation}
is a metric on the moduli space $\Mcal$, and the argument~$\tau$ reminds us of a dependence on the shape of~$T^2$.
One can also show that the equations (\ref{3.5}) are equivalent to the
Euler-Lagrange equations for $\phi^\a$ following from (\ref{3.18})  (cf.~\cite{11}). Finally, substituting (\ref{3.17}) into (\ref{3.6}),
we arrive at
\begin{equation}\label{3.20}
\bigl( \de_a^c\de_b^d-\sfrac12g_{ab}g^{cd}\bigr)\,\pa_c\phi^\a\,\pa_d\phi^\b\,G_{\a\b} \= 0\ ,
\end{equation}
which can also be obtained from (\ref{3.18}) by varying the metric $g^{}_{\Si_2}$. 
These are the Virasoro-type constraint equations.

\section{Examples}

\noindent
Here we briefly discuss examples of $d=10$ manifolds considered in the string literature.
The list is not complete and serves only illustrative purposes. Superstring theories in all these backgrounds can be
obtained from Yang-Mills theory via the adiabatic limit $\ve^2\to 0$ as discussed in the previous section.

\medskip

\noindent {\bf AdS$_4\times\C P^3$. } The background
\begin{equation}\label{4.1}
G/H\=\mbox{AdS}_4\times\C P^3 \= \frac{\mbox{SO(3,2)}}{\mbox{SO(3,1)}}\times\frac{\mbox{SU(4)}}{\mbox{U(3)}}
\end{equation}
is considered in the context of the AdS$_4/$CFT$_3$ correspondence relating the IIA string in the coset (\ref{4.1})
with $\Ncal{=}\,6$ super-Chern-Simons theory in three dimensions. 
Here $\C P^3$ is the standard complex projective space fibered over $S^4$ with $\C P^1$-fibres,
\begin{equation}\label{4.2}
\begin{CD}
\C P^3    @>{\C P^1}>>  S^4\ .
\end{CD}
\end{equation}
It has an integrable almost complex structure $\Jcal_+$ defining (1,0)-forms $\om^a$ on $\C P^3$ ($a=1,2,3$) via 
$\Jcal_+\om^a=\im\om^a$.

\medskip

\noindent {\bf AdS$_4\times\C P^3_{qK}$.} The background (\ref{4.1}) is not suitable for the consideration of heterotic
strings since the K\"ahler space $\C P^3$ has a U(3) holonomy. The situation is changed if one switches from the
integrable almost complex structure $\Jcal_+$ on $\C P^3$ to a non-integrable one $\Jcal_-$, which defines a 
quasi-K\"ahler space $\C P^3_{qK}$ isomorphic to $\C P^3$ as a smooth manifold.
The (1,0)-forms $\Theta^a$ with respect to $\Jcal_-$, obeying  $\Jcal_-\Theta^a=\im \Theta^a$, 
relate to the previous ones as follows~\cite{24},
\begin{equation}\label{4.4}
\Theta^1=\om^{\1}\ ,\quad\Theta^2=\om^{\2}\and \Theta^3=\om^3\ .
\end{equation}
The manifold $\C P^3_{qK}$, defined by $\Jcal_-$ and the (1,0)-forms (\ref{4.4}), has the structure group
U(2)$\,\subset\,$SU(3), and its almost complex structure $\Jcal_-$ is non-integrable due to torsion~\cite{24}. 
Let $\Lambda$ be the radius of $S^4$ and $R$ be the radius of $\C P^1$ from (\ref{4.2}). 
For $\Lambda^2=2R^2$ the space $\C P^3_{qK}$ is nearly K\"ahler and the torsion is totally antisymmetric.
Since the latter may then be identified with the $H$-field flux, such manifolds appear in heterotic string 
compactifications with fluxes (see e.g.~\cite{25,26,27} and references therein). 

\medskip

\noindent {\bf Resolved conifold.}   
The resolved conifold $\Ocal (-1)\oplus\Ocal (-1)\to \C P^1$ can be obtained in our approach
by considering a moduli space of flat connections on the punctured $T^2$ of the form
\begin{equation}\label{4.10}
G/H \= \R^{3,1}\times \frac{\mbox{SU}(5)}{\mbox{U}(4)} \= \R^{3,1}\times \C P^4
\end{equation}
and restricting to the non-singular {\it quintic threefold\/} in $ \C P^4$ 
(the zeros of a homogeneous quintic polynomial in the homogeneous $ \C P^4$ coordinates)~\cite{28}. 
The same trick can be employed in the approach of Shifman and Yung,
since $ \C P^N$ spaces are the standard moduli spaces of non-Abelian vortices (see reviews~\cite{4}-\cite{7}).

\medskip

\noindent {\bf Space $T^*S^3$}. One can always view the cotangent bundle  $T^*K$ of a Lie group $K$ as a Lie group. 
To this end, one performs a left trivialization (admitted by the parallelizability of $K$) and endows the resulting
trivial bundle $K\times (\Lie K)^*$ with the semi-direct product $K\ltimes (\Lie K)^*$ by using the coadjoint action of $K$ on
the space $(\Lie K)^*$ dual to $\Lie K$. In the case of $K=\;$SU(2) we can identify $su(2)^*$ with $su(2)$ and consider the
six-dimensional real group manifold SU(2)$\ltimes su(2)$, which is diffeomorphic to the deformed conifold $T^*S^3$.
Choosing a proper metric tensor $G_{\a\b}$ on this space, one can obtain string theory on 
$\R^{3,1}\times T^*S^3$ as the low-energy limit of Yang-Mills theory.

\medskip

\noindent {\bf Flat $d=10$ superspace}. 
For obtaining the Green-Schwarz superstring action (of type I, IIA or IIB) one should employ
supergroups $\tilde G$ instead of Lie groups $G$ which can be embedded in $\tilde G$ as bosonic subgroups, 
$G\subset\tilde G$, and the infrared limit of the corresponding supergroup gauge theories. 
This was demonstrated for superstrings in~\cite{11,13} and for supermembranes in~\cite{12}. Those papers
treated the moduli space of flat connections on the circle $S^1$ or on the disk $H^2$ with proper boundary conditions. 
Here instead we use the moduli space of super-Lie-algebra valued flat connections on the punctured $T^2$. 
This moduli space is a finite-dimensional supercoset space
\begin{equation}\label{4.11}
\tilde\Mcal =\tilde G/H\ ,
\end{equation}
and the analysis is simpler than in~\cite{13} where moduli spaces were loop supercosets. 
However, the derivation of the low-energy limit is so similar that we will not repeat it here 
and describe only the final results.

So, for superstrings moving in Minkowski space $\R^{9,1}$, one should extend the bosonic Lie group of translation $G=\R^{9,1}$ to the supergroup (cf.~\cite{29})
\begin{equation}\label{4.12}
\tilde G \=\frac{N{=}2\,\mbox{SUSY}}{\mbox{SO}(9,1)}\ ,
\end{equation}
which is a subgroup of the $N{=}2$ super Poincar\'e group in ten dimensions generated by translations and $N{=}2$
supersymmetry transformations. Coordinates on $\tilde G$ are $(X^\Delta) = (X^\a, \th^{Ap})$, where $X^\a$ with $\a =0,\ldots,9$
parametrize $\R^{9,1}$ and $\th^{Ap}$ with $A=1,\ldots,32$ and $p=1,2$ are the components of two Mayorana-Weyl spinors $\th^p$.
By considering Yang-Mills theory  on $M^4=\Si_2\times T^2$ with $\tilde G$ as the gauge group and taking the
adiabatic $\ve^2\to 0$ limit in (\ref{8}), we get a string moving in the moduli space $\tilde G$ of flat connections on
the punctured $T^2$. Its action functional reads
\begin{equation}\label{4.13}
S_0\ =\int_{\Si_2} \diff x^1 \diff x^2\,\sqrt{|\det g^{}_{\Si_2}|}\,g^{ab}\,\eta_{\a\b}\,\Pi^\a_a\,\Pi^\b_b\ ,
\end{equation}
where $\eta =(\eta_{\a\b})$ is the Minkowski metric on $\R^{9,1}$, and
\begin{equation}\label{4.14}
\Pi^\Delta_a=(\Pi^\a_a\ ,\ \Pi^{Ap}_a) \quad\with 
\Pi^\a_a=\pa_aX^\a-\im\de_{pq}\bar\th^p\ga^\a\pa_a\th^q \and \Pi^{Ap}_a=\pa_a\th^{Ap}
\end{equation}
are the components of one-forms $\Pi^\Delta= \diff x^a\Pi^\Delta_a$ on $\Si_2$ pulled back from one-forms $\diff X^\a$
and $\diff\th^{Ap}$ on $\tilde G$. Finally, $\ga^\a$ are $\ga$-matrices in $\R^{9,1}$ and $\bar\th^p:=(\th^p)^\top C$, 
where $C$ is the charge conjugation matrix.

The action  (\ref{4.13}) is not yet the full Green-Schwarz action, which needs an additional Wess-Zumino-type term~\cite{30}.
This term may also be obtained from supergroup gauge theory, by extending $\Si_2$ and $T^2$ to a Lorentzian 3-manifold 
$\Si_3$ with boundary $\pa\Si_3=\Si_2$ (as in~\cite{29}) and to a Riemannian 3-manifold $B^3$ with boundary $\pa B^3 = T^2$
(as in~\cite{13} for $H^2$). Then on $M^6=\Si_3\times B^3$ one can formulate the topological Yang-Mills term
\begin{equation}\label{4.15}
S_{W\! Z}\ =\int^{}_{\Si_3\times  B^3} f_{\Gamma\Delta\Lambda}\
\hat\Fcal^\Gamma\wedge\hat\Fcal^\Delta\wedge\hat\Fcal^\Lambda
\end{equation}
for a $\Lie\tilde G$-valued gauge field $\hat\Fcal$ on $M^6$, 
where the structure constants $f_{\Gamma \Delta \Lambda}$ are given in~\cite{29}. 
By the same calculations as in~\cite{13} one finds that in
the low-energy limit $\ve^2\to 0$ the action  (\ref{4.15}) reduces to a Wess-Zumino-type action functional~\cite{29,30} which
should be added to (\ref{4.13}) with a proper coefficient. Also, similarly to~\cite{13} one can show that the Kalb-Ramond
$B$-field appears from the topological term $\eta_{\alpha\beta} \Fcal^{\alpha}{\wedge}\Fcal^{\beta}$, 
whose integral in the adiabatic limit $\ve^2\to0$ becomes
\begin{equation}\label{4.15a}
\int_{M^4}^{} \diff^4x\, \ve^{ab} \ve^{ij}\langle\Fcal_{ai},\Fcal_{bj}\rangle \=
\int_{\Si_2} \diff x^1\diff x^2 \, \ve^{cd}\, B_{\a\b}\, \pa_c X^\a \pa_d X^\b\ ,
\end{equation}
where
\begin{equation}\label{4.15b}
B_{\a\b}\ = \int^{}_{T^2} \diff x^3\diff x^4 \,\ve^{ij}\langle\xi_{\a i},\xi_{\b j}\rangle
\end{equation}
are components of a two-form $\mathbb{B}=(B_{\a\b})$ on the moduli space (\ref{4.12}).

\medskip

\noindent {\bf AdS$_5\times S^5$.} The coset space
\begin{equation}\label{4.16}
G/H \= \mbox{AdS}_5\times S^5 \= \frac{\mbox{SO}(4,2)}{\mbox{SO}(4,1)} \times \frac{\mbox{SO}(6)}{\mbox{SO}(5)}
\end{equation}
is important in the AdS$_5$/CFT$_4$ correspondence between type IIB strings on this coset space and
$\Ncal{=}\,4$ super-Yang-Mills theory on the boundary $\R^{3,1}$ of AdS$_5$. The group 
$G=\mbox{SO}(4,2)\times\mbox{SO}(6)$ can be
embedded into the supergroup  $\tilde G=\;$PSU(2,2$|$4), and the supercoset $\tilde G/H$ with 
$H=\mbox{SO}(4,1)\times\mbox{SO}(5)$ 
is used for describing the superstring action~\cite{31}. 
Considering gauge theory with the supergroup $\tilde G=\,$PSU(2,2$|$4)
on $M^4=\Si_2\times T^2$, we get in the  $\ve^2\to 0$ limit the moduli space $\tilde G/H$ of flat connections on $T^2$
as the string target space. Both (\ref{4.13}) and (\ref{4.15}) will apply
with a proper choice of $G_{\alpha\beta}$ and $f_{\Gamma \Delta \Lambda}$ 
on $G/H$ and ${\tilde G}/H$, because in this limit the non-vanishing
components of $\Fcal$ (and  $\hat\Fcal$) are proportional to the pull-back 
\begin{equation}\label{4.17}
L^\Delta =(\diff X^M)L^\Delta_M \quad\to\quad\Pi^\Delta = (\diff x^a)\Pi^\Delta_a
\qquad\textrm{where}\quad \Pi^\Delta_a=(\pa_a X^M)L^\Delta_M\ ,
\end{equation}
and the index $\Delta$ runs over the coset parts of the generators of $psu(2,2|4)$~\cite{31}. The explicit form
of the superstring action (both kinetic and WZ terms) in terms of $\Pi^\Delta_a$ can be found in~\cite{31}. 
Similarly one can derive the full type IIA string action on AdS$_4\times\C P^3$ by considering supergroup gauge theory 
on $\Si_2\times T^2$ with $\tilde G=\;$OSp(2,2$|$6) and $H=\;$SO(3,1)$\times$U(3). 
Note that in (\ref{4.13}) one will have the metric $G_{\a\b}$ on the coset $G/H$ instead of $\eta_{\a\b}$.

\section{Conclusions}

\noindent
We have shown that the Yang-Mills action on the product of a two-dimensional Lorentzian manifold $\Si_2$ 
and a singly-punctured two-torus $T^2_p$, augmented by a topological term, flows to the Green-Schwarz 
superstring action on the worldsheet~$\Si_2$ in the infrared limit, when $T^2_p$ shrinks to a point.
Upon choosing a supergroup $\tilde G$ as the gauge group and picking a closed subgroup~$H\subset\tilde G$,
the string target space becomes the supercoset $\tilde G/H$ as the moduli space of flat Yang-Mills connections on~$T^2_p$.
We mainly focused on the bosonic part of the superstring action because we want to emphasize the fundamental possibility 
of receiving superstring sigma models in an infrared limit of corresponding suitable Yang-Mills theories. 
A lot of backgrounds, including PSU(2,2$|$4)/SO(4,1)$\times$SO(5) and OSp(2,2$|$6)/SO(3,1)$\times$U(3), 
may appear as moduli spaces of flat connections on $T^2_p$. 
Various other backgrounds can be obtained by generalizing the $T^2_p$ factor to a Riemann surface~$\tilde\Si_2$
with punctures or boundaries, whose moduli space of flat connections will depend on the geometry and boundary conditions.
In the infrared limit of gauge theory on $\Si_2\times \tilde\Si_2$, this moduli space becomes the target space
of a string sigma model on~$\Si_2$, promising a fresh perspective on the string vacuum landscape.
Clearly, the relation between Yang-Mills and string theories deserve further study.

\medskip

\noindent {\bf Acknowledgements}

\noindent 
This work was partially supported by the Deutsche Forschungsgemeinschaft grant LE 838/13.
This article is based upon work from COST Action MP1405 QSPACE,
supported by COST (European Cooperation in Science and Technology).

\end{document}